\documentclass[aps,prb,twocolumn,showpacs,floatfix,groupedaddress,superscriptaddress]{revtex4-2}
\pdfoutput=1
\usepackage[linkcolor=blue,colorlinks=true,breaklinks=true,citecolor=blue,urlcolor=blue]{hyperref}
\usepackage{graphicx,graphics} 
\usepackage{tabularx}
\usepackage{dcolumn}   
\usepackage{braket}
\usepackage[mathscr]{euscript}
\usepackage{amsbsy}
\usepackage{bm} 
\usepackage{amssymb} 
\usepackage[T1]{fontenc}
\usepackage{amsmath}
\usepackage{xcolor}
\usepackage[caption=false,labelformat=simple]{subfig}
\usepackage{bbold}

\makeatletter
\newcommand{\phantomlabelabovecaption}[2]{
	\protected@write\@auxout{}{
		\string\newlabel{#2}{
			{\number\numexpr\thefigure+1\relax#1}{\thepage}
			{\number\numexpr\thefigure+1\relax#1}{#2}{}
		}
	}
	\hypertarget{#2}{}
}
\makeatother

\begin{document}
\title{Commensurate moir\'e superlattices in anisotropically strained twisted bilayer graphene}
\author{Ayan Mondal}
\email{ayanmondal367@gmail.com}
\affiliation{Department of Physical Sciences, Indian Institute of Science Education and Research Kolkata\\ Mohanpur-741246, West Bengal, India}
\author{Cristina Joseph}
\affiliation{Department of Physical Sciences, Indian Institute of Science Education and Research Kolkata\\ Mohanpur-741246, West Bengal, India}
\author{Bheema Lingam Chittari}
\email{bheemalingam@iiserkol.ac.in}
\affiliation{Department of Physical Sciences, Indian Institute of Science Education and Research Kolkata\\ Mohanpur-741246, West Bengal, India}
\begin{abstract}
We investigate how anisotropic strain reorganizes commensurate moir\'e superlattices and electronic structure in twisted bilayer graphene (TBG) across a finite range of reference twist angles. Motivated by experiments showing robust moir\'e phenomenology under angular disorder and heterostrain (Kapfer et al., Science 381,677 (2023)), we construct commensurate strained supercells generated by a general anisotropic deformation of the top graphene layer of TBG. The results show that anisotropic strain does not generically destroy the electronic structure of nearby pristine moir\'e systems; rather, its effect depends sensitively on whether the strained commensurate geometry remains two dimensional or crosses over toward a quasi one dimensional regime. This provides a geometric perspective on the persistence of moir\'e electronic features over a finite window of twist angle and heterostrain. Within this framework, the allowed strained configurations naturally separate into tilted two dimensional moir\'e patterns and quasi one dimensional stripe like patterns. We find that several such strained two dimensional solutions occur near a given pristine twist angle, and that nearby solutions retain triangular like AA-region localization, comparable low energy bandwidths, and a low field Hofstadter spectrum close to the unstrained system. In contrast, quasi one dimensional strained configurations show stronger dimensional reduction, reduced Dirac point multiplicity, stripe like spatial localization, and stronger Hofstadter splitting.
\end{abstract}
\maketitle
\section{Introduction}\label{I}
The relative rotation between two graphene layers generates a long wavelength moir\'e superlattice that profoundly reshapes the Dirac spectrum, producing mini Brillouin zones and strongly hybridized electronic bands~\cite{Santos,Shallcross1,Shallcross2,Laissardière,Bistritzer}. At specific twist angles, most notably the so called magic angles, the Dirac velocity is strongly quenched, and nearly flat bands emerge at charge neutrality~\cite{Santos1,Carr1,Tarnopolsky}. This extreme sensitivity to a single geometric parameter has established TBG as a minimal and highly tunable platform for exploring strong correlation effects and topological phenomena in two dimensions~\cite{Andrei,Choi,Liu1,Zhang5,Cao1,Lake,Cao2,Po,Lee,Sharpe,Serlin}. Beyond twist angle control, external parameters provide additional tunability~\cite{Talkington,Sinha,Dutta,Carr,Chittari,Chebrolu,Padhi,Yankowitz1,Mondal}. Finite lattice strain is an inherent feature of experimentally realized TBG samples~\cite{Qiao,Mendoza,Wagner}. It is not merely a device imperfection but often a hidden control knob that determines the effective moir\'e potential experienced by electrons~\cite{Sanctis,Bi}. Experiments that directly reconstruct lattice deformations show that realistic samples host sizeable and spatially varying heterostrain~\cite{Kazmierczak,Kapfer}. This raises a critical question: does anisotropic strain destroy the exquisite moir\'e physics of TBG, or can it be harnessed as a tuning knob?  While the first magic angle was originally identified as a sharply defined twist angle near $1.08^\circ$, it is now well established that realistic devices exhibit substantial angular disorder and heterostrain. Twist angle variations of order $\pm 0.1^\circ$ and heterostrain in the range $0.1-0.7\%$ are generically present, and can stabilize flat band phenomenology and correlated states even away from the nominal magic angle~\cite{Kapfer,Yankowitz,Uri,Scheer,Lau}. As a result, the magic angle is more appropriately viewed as a finite window in the combined space of twist angle and strain, rather than a single pristine geometric value.
\par Structural relaxation further favours locally commensurate stacking configurations over perfectly incommensurate moir\'e patterns, leading to spatially heterogeneous electronic landscapes~\cite{Shi,Rakib,Meng}. Even at essentially fixed twist angle, changes in heterostrain can reorganize lattice relaxation and strongly modify the resulting moir\'e landscape~\cite{Escudero}. Theoretically, even a sub percent relative strain between the two layers generically reshapes the moir\'e Brillouin zone and shifts the relative Dirac point geometry~\cite{Bi,Khatibi,Zhang}. Beyond twist only symmetry considerations, strain can split or broaden van Hove singularities and strongly alter low energy band connectivity~\cite{Yan,Huder,Mesple,Kitt}. Heterostrain can therefore be exploited as an engineering tool to access flat band phenomenology away from standard conditions, including correlated electronic features at non magic angles~\cite{Manna,Li,Nakatsuji,Zhang1}. {More broadly, the interplay of twist and strain enables a wide family of moir\'e geometries beyond a simple stretched hexagon, including reduced dimensional stripe like structures reported in previous heterostrain studies, although certain strain configurations may also suppress flat band formation by strongly distorting the superlattice}~\cite{Escudero,Gao}. The emerging strain twistronics framework establishes strain as a central control parameter that must be incorporated for any realistic and predictive description of TBG~\cite{Hou,Kögl}.
\par In the presence of a perpendicular magnetic field, magnetotransport and spectroscopic studies have established that Hofstadter physics in twisted bilayer graphene is highly sensitive to lattice deformation and heterostrain \cite{Finney, Wang}. Experiments have shown that nominally identical twist angles can exhibit markedly different Landau fan structures, Hall sequences, and Hofstadter butterflies, with strain identified as the dominant source of this variability \cite{Finney, Lu}. In particular, heterostrain has been observed to broaden, split, or reorganize Hofstadter subbands, modify Chern number assignments, and induce asymmetries in the quantum Hall response \cite{Wang}. Theoretically, strain is found to distort the moir\'e Brillouin zone and alter band connectivity, leading to substantial changes in the fractal spectrum even at fixed magnetic flux \cite{Lu, Zhang2}. However, a systematic understanding of how general anisotropic strain reorganizes commensurate moir\'e superlattices across twist angles remains incomplete. In particular, it is unclear whether anisotropic strain necessarily destroys moir\'e electronic features, or whether some strained geometries preserve two dimensional moir\'e physics while others cross over to a quasi one dimensional stripe like regime. To address this question, we construct commensurate strained supercells generated by a general anisotropic deformation of the top graphene layer and analyze their bandwidths, spatial localization, Dirac point structure, and Hofstadter spectra. The paper is organized as follows. Section~\ref{model} describes the anisotropically strained commensurate TBG model, local integer search protocol and the tight binding Hamiltonian. Section~\ref{results} presents the results: commensurate solution maps near \(\pm6.008^\circ\) (Sec.~\ref{6degree}), extension to other twist angles (Sec.~\ref{other_angles}), low energy band structures (Sec.~\ref{band}), spatially projected density of states (Sec.~\ref{SPDOS}), Hofstadter spectra (Sec.~\ref{Butterfly}), and weakly strained solutions near the magic angle (Sec.~\ref{magic}). Section~\ref{conclu} summarizes our conclusions.
\section{Model} \label{model}
\subsection{Anisotropic strain}
\par {\it Lattice parametrization: }
\begin{figure}
    \centering
    \includegraphics[width=0.5\columnwidth]{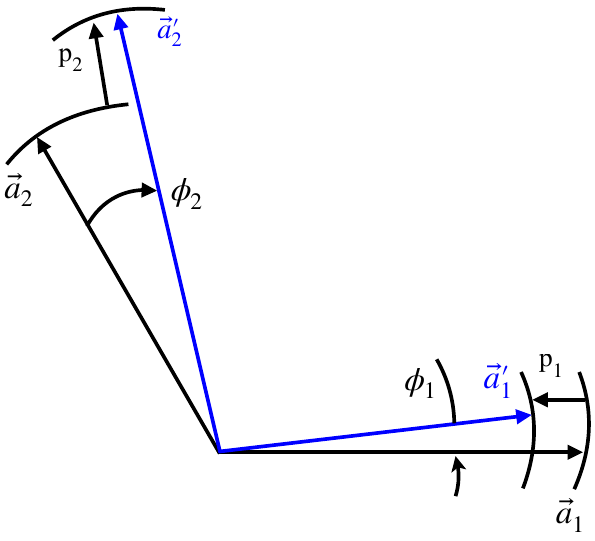}
    \caption{Schematic diagram illustrating a general anisotropic strain configuration used in this work. The lattice vectors of the unstrained bottom graphene layer are $\bm{a_1}$ and $\bm{a_2}$, shown in black. The lattice vectors of the top layer under anisotropic strain are denoted by $\bm{a_1}^\prime$ and $\bm{a_2}^\prime$, shown in blue. The vector $\bm{a_1}^\prime$ is obtained by compressing $\bm{a_1}$ by a factor $\textrm{p}_1$ and rotating it anticlockwise by an angle $\phi_1$, while $\bm{a_2}^\prime$ is obtained by elongating $\bm{a_2}$ by a factor $\textrm{p}_2$ and rotating it clockwise by an angle $\phi_2$.}
    \label{fig:fig1}
\end{figure}
The geometry of anisotropic strain applied to the top graphene layer of TBG relative to the unstrained bottom layer is illustrated in Fig.~\ref{fig:fig1}. We first describe the lattice parametrization of strain and commensuration. The pristine graphene has lattice vectors $\bm{a}_1$ $= a(1,0)$ and $\bm{a}_2$ $= a(-\frac{1}{2}, \frac{\sqrt{3}}{2})$ where $a =2.46 $ \AA~ is the lattice constant of graphene. The strained lattice vectors $\bm{a}_1^\prime$($\bm{a}_2^\prime$) are obtained by applying a scaling $\textrm{p}_1$($\textrm{p}_2$) and a rotation $\phi_1$($\phi_2$) to $\bm{a}_1$($\bm{a}_2$). Here, $\textrm{p}_i >1 (<1)$ corresponds to elongation (compression) of the corresponding lattice vector, while $\phi_i$ denotes its rotation relative to the bottom layer lattice vector. With the help of extended Wood’s notation~\cite{Artaud}, the strained lattice vectors of the top layer can be written in terms of the strain parameters as a linear transformation of the bottom layer lattice vectors:
\begin{equation}
    \begin{pmatrix}
  \bm{a}_1^\prime\\ 
  \bm{a}_2^\prime
\end{pmatrix}=\begin{pmatrix}
  \textrm{p}_1(cos\phi_1+\frac{sin\phi_1}{\sqrt{3}}) & \frac{2\textrm{p}_1}{\sqrt{3}}sin\phi_1 \\ 
  -\frac{2\textrm{p}_2}{\sqrt{3}}sin\phi_2 & \textrm{p}_2(cos\phi_2-\frac{sin\phi_2}{\sqrt{3}})
\end{pmatrix} \begin{pmatrix}
  \bm{a}_1 \\ 
  \bm{a}_2
\end{pmatrix}
\label{eq:eq1}
\end{equation}
\par {\it Commensuration conditions: } A commensurate moir\'e pattern is formed when integer multiples of the lattice vectors of the two layers coincide, allowing a common supercell to be defined. The lattice vectors of moir\'e unit cell ($\bm{A}_1$, $\bm{A}_2$) will then be related to the lattice vectors of each layer as 
\begin{equation}
    \begin{pmatrix}
  \bm{A}_1\\ 
  \bm{A}_2
\end{pmatrix}=\begin{pmatrix}
  i & j\\ 
  k & l
\end{pmatrix} \begin{pmatrix}
  \bm{a}_1^\prime \\ 
  \bm{a}_2^\prime 
\end{pmatrix}=\begin{pmatrix}
  m & n\\ 
  q & r
\end{pmatrix}\begin{pmatrix}
  \bm{a}_1 \\ 
  \bm{a}_2 
\end{pmatrix}
\label{eq:eq2}
\end{equation}
where $i,j,k,l,m,n,q,r$ are non zero integer numbers.
From Eq.~\ref{eq:eq2}, the lattice vectors of two layers in terms of the eight integers are related as \cite{Escudero1}:
\begin{equation}
    \begin{pmatrix}
  \bm{a}_1^\prime \\ 
  \bm{a}_2^\prime
\end{pmatrix}= \frac{1}{il-jk} \begin{pmatrix}
  lm-jq & ln-jr\\ 
  -km+iq & -kn+ir
\end{pmatrix} \begin{pmatrix}
  \bm{a}_1 \\ 
  \bm{a}_2
\end{pmatrix}
\label{eq:eq3}
\end{equation}
By equating Eqs.~(\ref{eq:eq1}) and (\ref{eq:eq3}), each set of eight integer commensurate solutions determines the lattice vector parameters \(\textrm{p}_1,\textrm{p}_2,\phi_1,\phi_2\).

\par { {\it Physical strain parameters: } The Cartesian deformation matrix that maps the unstrained bottom layer primitive vectors to the strained top layer primitive vectors is given by
\begin{equation}
 S = (\bm{a}'_1\ \bm{a}'_2)(\bm{a}_1\ \bm{a}_2)^{-1}.
 \label{eq:eq4}
\end{equation}
We decompose the deformation matrix $S$ into physical strain parameters via polar decomposition~\cite{Benschop}. Since $S$ represents a single homogeneous deformation of the top layer, it can be written as $S = W P$, where $W$ is a rigid rotation matrix and $P$ is a symmetric positive definite stretch matrix. The finite strain tensor is then $\epsilon = P - I$. The explicit mapping from the eight integers to \(\textrm{p}_1,\textrm{p}_2,\phi_1,\phi_2\), together with the extraction of the polar decomposition strain parameters, is given in Appendix~\ref{appA}. Different familiar strain types are contained within this homogeneous deformation framework. Isotropic or biaxial strain corresponds to an equal stretching of all lattice directions, whereas a general anisotropic strain stretches different directions by different amounts. A uniaxial like deformation is obtained when the stretching is dominant along one direction. If the stretching axes are not aligned with the graphene lattice axes, the same homogeneous deformation appears as a combination of normal strain and shear in the graphene coordinate frame. Therefore, configurations with \(\phi_1\neq\phi_2\) correspond to a single homogeneous anisotropic strain field rather than independent rotations of the two primitive lattice vectors.}
\subsection{Strained commensurate solution search} \label{search_moire}
Strain is ubiquitous in graphene samples used to create TBG. While the magnitude of this strain can reach up to $0.9 \%$ \cite{Ouyang}, even this small amount corresponds to a large moir\'e scale deformation when the system is near the magic angle $(\sim 1.085^\circ)$. Any deformation applied at the graphene lattice scale is significantly amplified within the moir\'e superlattice by a factor proportional to $1/\theta$ (where $\theta$ is the twist angle in radians). This powerful scaling relationship implies that the same effective moir\'e deformation can be achieved at a larger twist angle by applying a proportionally larger heterostrain. { To make this comparison precise, we define a dimensionless moir\'e scale deformation parameter
\begin{equation}
\mathcal D_M=
\frac{\varepsilon_\textrm{p}}{\rm rad(\theta_0)},
\label{eq:moire_deformation}
\end{equation}
where $\varepsilon_\textrm{p}=\max\left(|\textrm{p}_1-1|,|\textrm{p}_2-1|\right)$, \(\textrm{p}_1\) and \(\textrm{p}_2\) are the lattice vector scaling factors of the strained top layer of TBG and \(\rm rad (\theta_0)\) is the pristine reference twist angle expressed in radians. The quantity \(\mathcal D_M\) measures the microscopic lattice vector strain relative to the pristine moir\'e wave vector scale. For a strain magnitude of $\varepsilon_\textrm{p} = \frac{0.9}{100}\simeq0.9\%$ in TBG with $\theta_0 = 1.084^\circ$, we obtain
\begin{equation}
\mathcal D_M = \frac{0.009}{\mathrm{rad}(1.084^\circ)}\simeq 0.47,
\end{equation}
corresponding to approximately \(47\%\) moir\'e scale deformation. At a twist angle $\theta_0 = 6.008^\circ$, maintaining the same cutoff $\mathcal D_M \simeq 0.47$ allows microscopic lattice vector strains up to approximately $4.92\%$. Since a unique twist angle is generally not defined once \(\phi_1\neq\phi_2\), \(\theta_0\) in Eq.~\eqref{eq:moire_deformation} always denotes the pristine reference angle around which the commensurate search is performed, not the twist angle of the strained configuration.} 

{ For an unstrained commensurate twist angle \(\theta_0\) specified by two integers \((n_0,m_0)\), with \(m_0>n_0\)~\cite{Shallcross}, the corresponding eight integer representation is $ (i,j,k,l,m,n,q,r) = (n_0+m_0,\ n_0,\ m_0,\ n_0+m_0,\ n_0+m_0,\ m_0,\ n_0,\ n_0+m_0). $ To obtain strained commensurate solutions near this pristine reference angle, we search over the eight integers within the local window $ n_{\rm min}\le i,j,k,l,m,n,q,r\le n_{\rm max},$ where \(n_{\rm min}=n_0\) and \(n_{\rm max}=n_0+m_0\). This window is chosen around the unstrained commensurate representation so as to sample strained configurations continuously connected to the reference structure, while excluding unrelated commensurate families associated with different pristine twist angles. 

The integer search can also generate higher order commensurate cells. To exclude higher order repeated commensurate cells, we impose a first order beating condition \cite{Zeller}. Defining
\begin{align}
    N_1=&\sqrt{(i-m)^2+(j-n)^2-(i-m)(j-n)}, \nonumber \\
    N_2=&\sqrt{(k-q)^2+(l-r)^2-(k-q)(l-r)},
\end{align}
where the quadratic form is the graphene lattice metric, we retain only solutions satisfying $ N_1N_2=1 .$ This selects the minimal moir\'e supercell associated with the relative deformation. We further restrict the solutions by imposing \(\mathcal D_M\le 0.5\), together with an angular cutoff on the apparent rotation angles \(\phi_1\) and \(\phi_2\), in order to exclude strongly deformed configurations.} { The larger shear dominated deformations near  \(6.008^\circ\) should be viewed as controlled commensurate model geometries, whereas the solution near magic angle remains in the experimentally relevant small strain regime.}
\subsection{Tight binding Hamiltonian}
\par We have considered a tight binding Hamiltonian for all subsequent calculations. The Hamiltonian is given by
\begin{align}
 H=&- \sum_{\substack{i, j }}t(r_i-r_j)(\hat{c}_{i}^{\dagger}\hat{c}_{j}+H.c.) 
 \label{eq:eq5}
\end{align}
 Here $t$ denotes the hopping integral between two sites $i$ (position $\bm{r}_i$), $j$ (position $\bm{r}_j$) and is a function of their atomic distance. The operators $\hat{c}_{i}^{\dagger}$ and $\hat{c}_{i}$ create and annihilate a $p_z$ electron at site $i$, respectively. H.c. denotes the Hermitian conjugate term. We have taken Slater-Koster type \cite{Slater} formula for hopping integral: \cite{Uryu, Laissardi, Moon}  
\begin{align}
    -t(d) = &V_{pp\pi}(d)(1-\frac{\Vec{d}.\hat{e_z}}{d})^2+V_{pp\sigma}(d)(\frac{\Vec{d}.\hat{e_z}}{d})^2 \label{eq:eq5}\\
    &V_{pp\pi}(d) = V_{pp\pi}^0exp(-\frac{d-a_0}{\delta})\\
    &V_{pp\sigma}(d) = V_{pp\sigma}^0exp(-\frac{d-d_0}{\delta})
\end{align}
where $d = r_i-r_j$ is the distance between two atoms at sites i, j and $\hat{e_z}$ is the unit vector along the z axis. $V_{pp \pi} ^0= -2.7 $eV is the hopping integral between two nearest neighbour atoms with distance $a_0 = a/\sqrt{3}\sim 1.42$ \AA~. The decay length is chosen to be $\delta = 0.184a$ so that all the significant contributions to the hopping energy is included. $V_{pp \sigma}^0 = -0.48$eV is the hopping integral between two nearest vertically aligned atoms which are separated by interlayer spacing $d_0=3.34$ \AA~. These choices ensure a realistic decay of hopping amplitudes and accurately reproduce the electronic structure of twisted bilayer graphene \cite{Sinha, Mondal}.

In realistic samples, atomic positions relax to minimize the combined intralayer elastic energy and interlayer stacking (adhesion) energy, which in twisted bilayer graphene generally reduces the area of AA stacking and enhances AB/BA domains \cite{Nam}. In the presence of anisotropic strain, relaxation can partially redistribute the local distortion field and thereby reduce the effective local registry mismatch \cite{Ouyang}. Such structural rearrangements may influence higher energy states, which correspond to electronic modes at domain walls \cite{Nguyen, Timmel}. { In this work, we include in plane relaxation by minimizing the total energy consisting of the intralayer elastic energy and the interlayer stacking energy, as detailed in Ref.~\cite{Nam}.}
\section{Results and Discussions}\label{results}
\subsection{Commensurate moir\'e patterns}\label{6degree}
\begin{figure}
    \includegraphics[width=\linewidth]{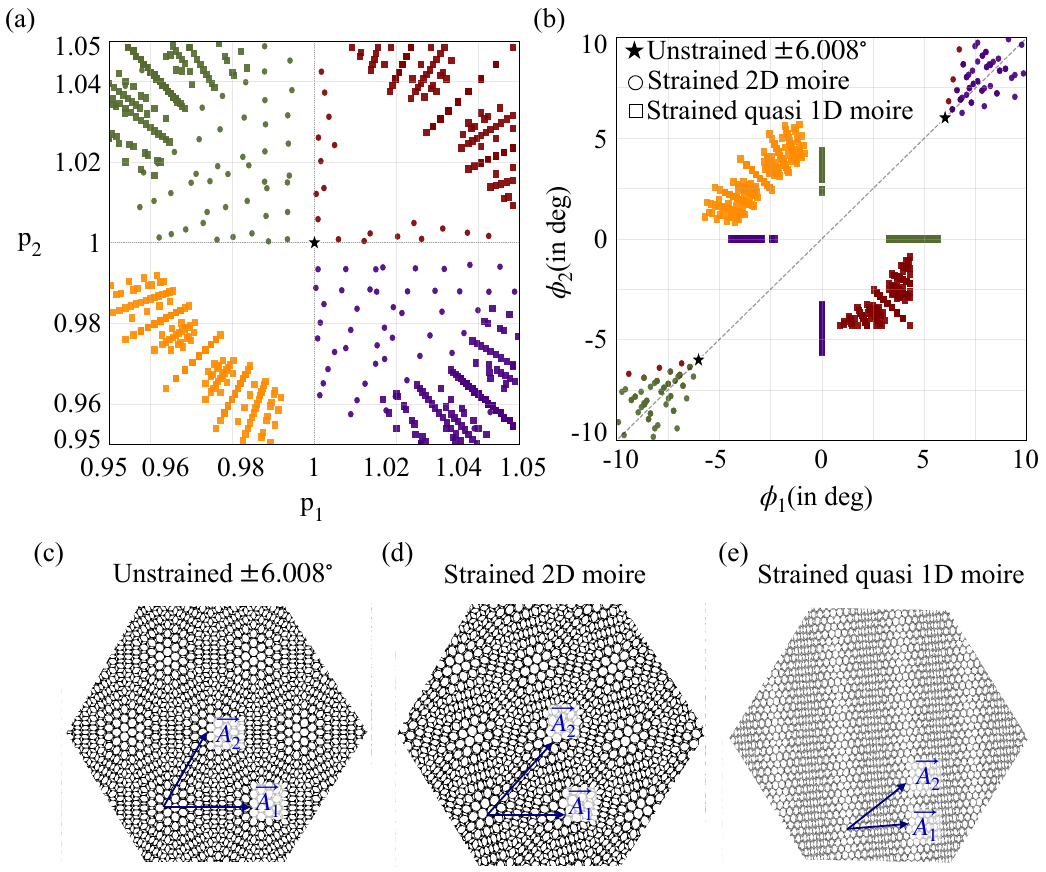}
    \caption{Phase space representation of strained commensurate moir\'e solutions around the pristine reference twist angles \(\pm6.008^\circ\) in the (a) \((\textrm{p}_1, \textrm{p}_2)\) and (b) \((\phi_1,\phi_2)\) parameter spaces. Here \(\textrm{p}_i\) denotes the scaling of the \(i\)-th strained primitive vector, while \(\phi_i\) denotes its apparent rotation relative to the corresponding bottom layer primitive vector. The pristine unstrained reference configurations are marked by stars. Circles and squares denote tilted two dimensional and quasi one dimensional solutions, respectively. The moir\'e patterns for (c) unstrained twisted bilayer graphene at a twist angle of $\pm 6.008^\circ$, and for commensurate anisotropic strain solutions producing (d) a two dimensional (2D) distorted moir\'e lattice and (e) a quasi one dimensional (quasi 1D) moir\'e lattice. The corresponding moir\'e lattice vectors $(\vec{A}_1, \vec{A}_2)$ are shown in blue.}
    \label{fig:fig2}
\end{figure}
\par { In this section, we systematically explore anisotropically strained commensurate configurations in the vicinity of an unstrained twist angle $6.008^\circ$,  a computationally accessible commensurate reference whose strained supercells remain small enough for systematic calculations.} The 8-integer set ({\it i, j, k, l, m, n, q, r}) corresponding to the unstrained commensurate configuration at a twist angle of $6.008^\circ (n_0,m_0=5,6)$ is $(11,5,6,11,11,6,5,11)$, which gives $\textrm{p}_1=\textrm{p}_2=1$ and $\phi_1=\phi_2=6.008^\circ$. The corresponding configuration for the opposite twist, $-6.008^\circ$, is $(11,6,5,11,11,5,6,11)$. { To identify strained commensurate solutions connected to the \(\pm6.008^\circ\) reference structures, we searched all eight integers in the range \(5 (n_0) \le i,j,k,l,m,n,q,r\le 11(n_0+m_0)\), retaining solutions with \(\mathcal D_M\le0.5\) and \(|\phi_1|,|\phi_2|\le10^\circ\). These cutoffs define a local search window around the \(\pm6.008^\circ\) branch. The angular cutoff excludes strongly rotated configurations, while the integer range isolates the nearby commensurate family.} Since each configuration can be specified either by an eight integer tuple or by the four physical parameters $(\textrm{p}_1,\textrm{p}_2,\phi_1,\phi_2)$, direct visualization of the solution space is nontrivial. We therefore represent all the commensurate solutions in the reduced $(\textrm{p}_1, \textrm{p}_2)$ and ($\phi_1, \phi_2$) parameter spaces, as shown in Fig.~\ref{fig:fig2}. The pristine, unstrained configuration is marked by a star. To better visualize the distribution of solutions, we use four colors to label the four quadrants in the $(\textrm{p}_1,\textrm{p}_2)$ space [ ~$\textrm{p}_1~\&~\textrm{p}_2 > 1$ (red), $\textrm{p}_1< 1~\&~\textrm{p}_2 > 1$ (green), $\textrm{p}_1~\&~\textrm{p}_2 < 1$ (orange) and $\textrm{p}_1> 1~\&~\textrm{p}_2 < 1$ (violet)] (see Fig.~\ref{fig:fig2}(a)). These correspond to the four possible combinations of compression and elongation of the two strained lattice vectors. The same color is assigned to the corresponding solution in the ($\phi_1,\phi_2$) space (see Fig.~\ref{fig:fig2}(b)). This allows a direct comparison between the strain and rotation parameterizations. This color coding is only a visualization aid. We classify the moiré patterns into two distinct types, tilted two dimensional (2D) and quasi one dimensional (quasi 1D), and represent them with circles and squares, respectively. 
\begin{table*}[t]
{ \caption{Physical strain parameters for representative commensurate solutions. The explicit extraction of the polar decomposition parameters of the deformation matrix \(S\) is described in Appendix~\ref{appA}. The rotation angle \(\xi\), obtained from \(W\), gives the overall rigid rotation of the top layer. The principal stretches \(d_1,d_2\), principal strains \(\epsilon_1,\epsilon_2\), strain anisotropy \(\kappa\), principal strain direction \(\psi\), and moir\'e dimensionality parameter \(\eta\) are all derived from the same deformation matrix \(S\).}
\begin{ruledtabular}
\begin{tabular}{c c c c c c c c c}
Integers & Type & \(\xi\) & \(d_1\) & \(d_2\) &
\(\epsilon_1,\epsilon_2\) & \(\kappa\) & \(\psi\) & \(\eta\) \\
\hline
\((11,5,6,11,11,6,5,11)\)
& Unstrained $6.008^\circ$
& \(6.008^\circ\)
& \(1.0\)
& \(1.0\)
& \(0\%,0\%\)
& \(1.0\)
& \(90.0^\circ\)
& \(1.0\)
\\
\((9,7,5,11,9,6,6,11)\)
& 2D tilted near $6.008^\circ$
& \(-7.31^\circ\)
& \(1.0163\)
& \(0.9686\)
& \(+1.63\%,-3.14\%\)
& \(1.049\)
& \(129.10^\circ\)
& \(0.684\)
\\
\((10,5,9,10,11,6,10,11)\)
& quasi 1D near $6.008^\circ$
& \(1.14^\circ\)
& \(1.1132\)
& \(0.9963\)
& \(+11.32\%,-0.37\%\)
& \(1.117\)
& \(48.90^\circ\)
& \(1.66\times10^{-15}\)
\\
\((31,59,59,30,30,59,59,31)\)
& 2D near magic angle
& \(1.13^\circ\)
& \(1.00045\)
& \(0.99955\)
& \(+0.045\%,-0.045\%\)
& \(1.001\)
& \(14.43^\circ\)
& \(0.955\)
\end{tabular}
\label{table1}
\end{ruledtabular}}
\end{table*}
\begin{table*}[t]
{ \caption{
Integer search windows used for the commensurate solution sets shown in Fig.~\ref{fig:fig3}. For each pristine reference angle \(\theta_0\), the search is performed locally around the corresponding unstrained eight integer solution within the stated strain and rotation cutoffs.}
\begin{ruledtabular}
\begin{tabular}{c c c c c}
Reference angle \(\theta_0\) &
\((n_0,m_0)\) &
Pristine integer tuple &
Integer window &
Cutoffs \((\varepsilon_{\rm p},|\phi_1|~\&~ |\phi_2|)\) \\
\hline
\( 4.408^\circ\) & \((7,8)\) &
\((15,7,8,15,15,8,7,15)\) &
\(7\le i,j,k,l,m,n,q,r\le 15\) &
\((3.8\%,8^\circ)\)\\

\( 5.085^\circ\) & \((6,7)\) &
\((13,6,7,13,13,7,6,13)\) &
\(6\le i,j,k,l,m,n,q,r\le 13\) &
\((4\%,8^\circ)\)\\

\( 7.340^\circ\) & \((4,5)\) &
\((9,4,5,9,9,5,4,9)\) &
\(4\le i,j,k,l,m,n,q,r\le 9\) &
\((6\%,10^\circ)\)\\

\( 9.430^\circ\) & \((3,4)\) &
\((7,3,4,7,7,4,3,7)\) &
\(3\le i,j,k,l,m,n,q,r\le 7\) &
\((8\%,15^\circ)\)
\end{tabular}
\label{table2}
\end{ruledtabular}}
\end{table*}
{ The two dimensional versus quasi one dimensional character of the moir\'e pattern is controlled by the relative deformation matrix \cite{Sinner}
\begin{equation}
D_{\rm rel}=S-I.
\end{equation}
Here \(S\) is the deformation matrix defined in Eq.~\ref{eq:eq4}. We define
\begin{equation}
\eta =
\sqrt{
\frac{
\lambda_{\min}
}{
\lambda_{\max}
}
},
\end{equation}
where \(\lambda_{\min}\) and \(\lambda_{\max}\) are the smaller and larger eigenvalues of \(D_{\rm rel}^{T}D_{\rm rel}\), respectively. If \(\eta\sim 1\), the relative stacking varies in two independent directions and the moir\'e pattern remains two dimensional. If \(\eta\ll1\), the matrix \(S-I\) is nearly rank one, implying the existence of an approximate invariant direction \(\mathbf v\) satisfying \((S-I)\mathbf v\simeq0\). The stacking then varies predominantly in the perpendicular direction, producing a stripe like quasi one dimensional moir\'e pattern.} Strain alone does not uniquely determine whether a configuration is 2D or quasi 1D. As shown in  Fig.~\ref{fig:fig2}(b), the distinction is controlled by the relative direction of rotation of the strained lattice vectors. When both vectors rotate in the same direction (clockwise or counterclockwise), the resulting moir\'e pattern remains slightly tilted two dimensional. In contrast, when the vectors rotate in opposite directions, or when only one of them rotates,  the system transitions to a quasi one dimensional pattern. We show in Fig.~\ref{fig:fig2}(c–e) how the moir\'e pattern evolves under anisotropic strain. The strained two dimensional moiré configurations retain a tilted triangular moiré lattice with displaced AA regions, whereas the quasi one dimensional moiré configurations exhibit a stripe like stacking modulation. All strained commensurate solutions break the $C_3$ 
rotational symmetry of pristine twisted bilayer graphene.\\
\begin{figure*}
    \centering
    \includegraphics[width=0.95\linewidth]{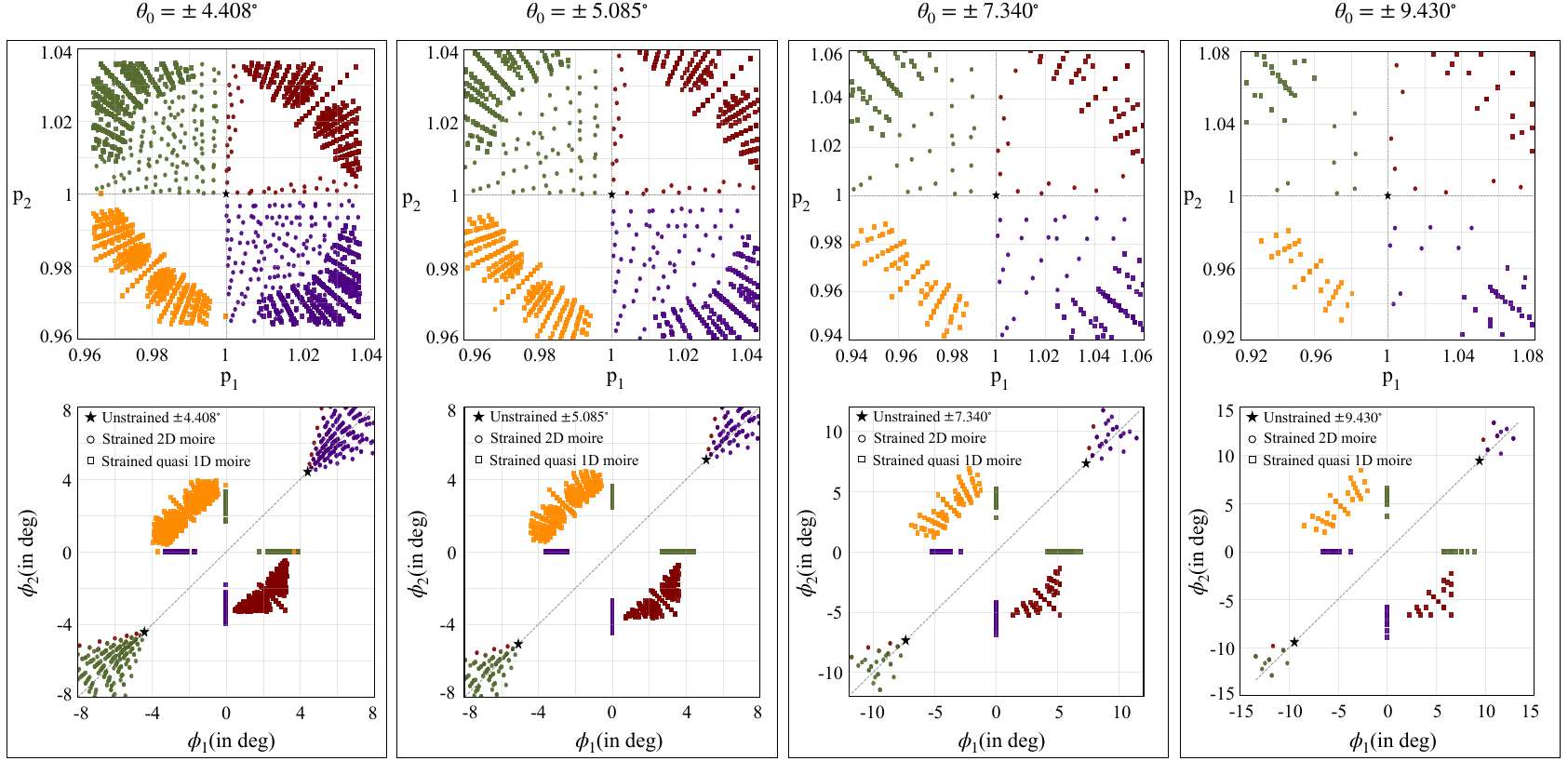}
    \caption{Commensurate anisotropically strained solutions around several pristine reference twist angles, shown in the \((\textrm{p}_1,\textrm{p}_2)\) and \((\phi_1,\phi_2)\) parameter spaces. Here \(\textrm{p}_i\) denotes the scaling of the \(i\)th strained primitive vector, and \(\phi_i\) denotes its apparent rotation relative to the corresponding bottom layer primitive vector. { For each reference angle \(\theta_0\), the search is performed within the local integer window listed in Table~\ref{table2}, retaining solutions within the fixed moir\'e deformation cutoff \(\mathcal D_M\le0.5\) and an angular cutoff. Circles and squares denote tilted two dimensional and quasi one dimensional solutions, respectively, classified using the dimensionality parameter \(\eta\).} }

    \label{fig:fig3}
\end{figure*}
{ Representative commensurate solutions near the pristine reference twist angles \(\pm 6.008^\circ\) are listed in Table~\ref{table1}, together with their physical strain parameters obtained from the polar decomposition analysis described in Appendix~\ref{appA}. The quasi one dimensional solution, with principal strain \(\epsilon_1 = +11.32\%\), should be viewed as an extreme controlled commensurate geometry that clearly illustrates the crossover toward quasi one dimensional behaviour, rather than as a typical passive heterostrain configuration occurring in experiments. In the next section, we show that the same geometric classification persists for commensurate solutions around several other pristine reference twist angles.}

\subsection{Strained commensurate moir\'e around various twist angles}\label{other_angles}   
{ In this section, we verify that the separation between tilted two dimensional and quasi one dimensional strained moir\'e patterns is not specific to the \(\pm 6.008^\circ\) reference structures. To this end, we repeat the commensurate search around several other pristine reference twist angles: \(\pm 4.408^\circ\), \(\pm 5.085^\circ\), \(\pm 7.340^\circ\), and \(\pm 9.430^\circ\). As shown in Fig.~\ref{fig:fig3}, the same qualitative organization appears across these angles when the solutions are compared at a fixed effective moir\'e deformation cutoff (\(\mathcal D_M \le 0.5\)) and within the angular cutoffs listed in Table~\ref{table2}.} As the twist angle increases, the number of admissible commensurate solutions decreases progressively, as expected from the smaller number of nearby integer solutions within the corresponding local search windows. When we consider all reference angles studied here, two distinct classes of strained commensurate solutions emerge. This observation indicates that strained commensurate moir\'e supercells of both classes are generically accessible around any unstrained commensurate twist angle. When the two strained top layer primitive vectors acquire apparent rotations in the same direction relative to the bottom layer, the resulting structures remain tilted two dimensional moir\'e patterns. In contrast, when the two apparent rotations occur in opposite directions, the solutions cross over to quasi one dimensional stripe like moir\'e patterns. If we examine the underlying hexagonal lattice geometry, this dichotomy follows naturally and remains consistent across all twist angles considered, { as discussed in Appendix~\ref{app:apparent_rotation}. As we will show in the following sections, the detailed electronic spectra presented later for the \(\pm 6.008^\circ\) structures thus serve as representative examples of this geometric classification, whereas the classification itself is more general and applies across a wide range of twist angles.} 
\begin{figure}
    \centering
    \includegraphics[width=0.95\linewidth]{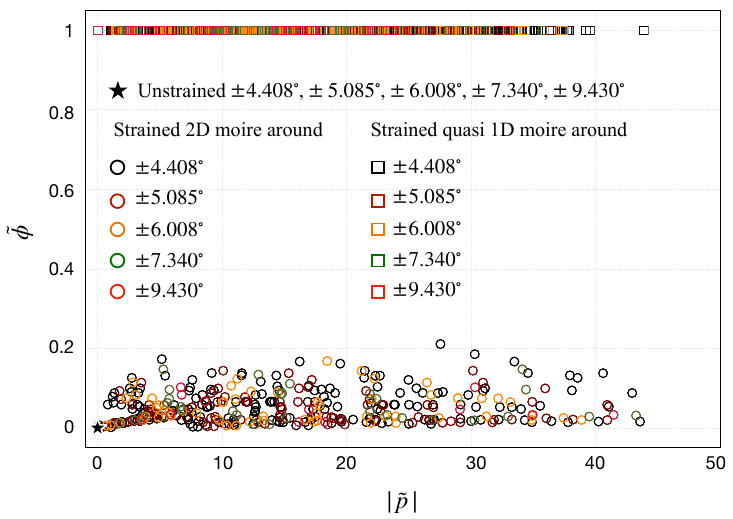}
    \caption{Commensurate moir\'e solutions under anisotropic strain around several pristine reference twist angles, shown as a function of the normalized rotational anisotropy
$
\widetilde{\phi}
$
and { the absolute moir\'e scale strain anisotropy
$
|\widetilde{\textrm{p}}|.
$ The plotted reference angles are
\(\pm4.408^\circ\), \(\pm5.085^\circ\), \(\pm6.008^\circ\), \(\pm7.340^\circ\), and \(\pm9.430^\circ\). }}
    \label{fig:fig4}
\end{figure}
\par This distinction is particularly evident in Fig.~\ref{fig:fig4}, where solutions from different reference angles are compared at fixed effective moir\'e deformation rather than fixed microscopic strain. We represent the solutions using the normalized rotational anisotropy $ \widetilde{\phi} = \frac{|\phi_1-\phi_2|}{|\phi_1|+|\phi_2|} $ and the absolute moir\'e scale strain anisotropy $ |\widetilde{\textrm{p}}| = \left| \frac{|\textrm{p}_1|-|\textrm{p}_2|}{\mathrm{rad}(\theta_0)} \right|. $ { Here \(\theta_0\) denotes the pristine reference twist angle of the unstrained commensurate structure around which the integer search is performed. It should not be interpreted as the twist angle of the strained configuration, since a unique twist angle is generally not defined when \(\phi_1\neq\phi_2\).} The two moir\'e regimes are clearly separated in this representation: \(\widetilde{\phi}=1\) for quasi one dimensional configurations, indicating opposite apparent rotations of the two strained primitive vectors, whereas \(\widetilde{\phi}<1\) for tilted two dimensional moir\'e patterns. Beyond the deformation window considered here, at larger strain values, this clean separation between quasi one dimensional and two dimensional solutions gradually breaks down. The quoted moir\'e deformation should therefore be understood as a distortion of the emergent interference pattern, not as an atomic scale strain. Since the moir\'e pattern is a geometric interference structure rather than a physical lattice of atoms, large moir\'e scale distortions can arise from comparatively small microscopic strain.
\subsection{Electronic band structure under anisotropic strain}\label{band}
{ In the following sections, we focus on representative structures near \(\pm6.008^\circ\), whose physical strain parameters are listed in Table~\ref{table1} and whose commensurate moir\'e supercell geometries are given in Table~\ref{table3}. These cells are small enough for detailed tight binding calculations. These results are intended to illustrate the electronic consequences of the three geometric classes identified above.} 
\begin{figure*}
    \centering
    \includegraphics[width=0.9\textwidth]{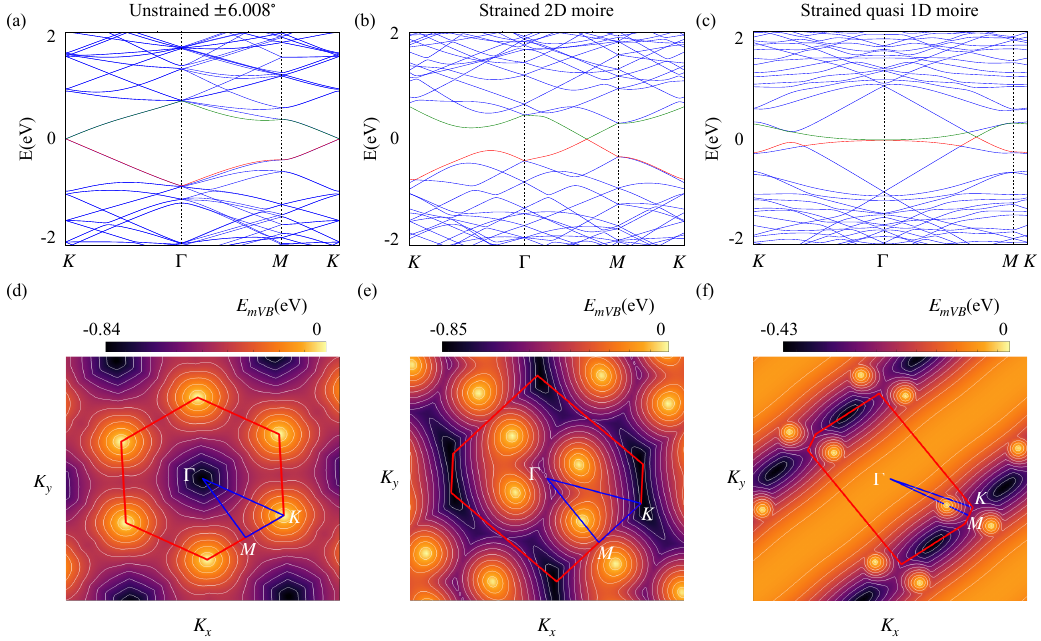}
    \caption{ Band structures along the high symmetry path \(K-\Gamma-M-K\) (top row) and corresponding surface plots of the highest valence band in the moir\'e Brillouin zone (bottom row) for three representative commensurate structures as given in Table~\ref{table3}. Panels (a,d) show the unstrained reference structure, panels (b,e) show the tilted two dimensional strained structure and panels (c,f) show the quasi one dimensional strained structure. In the unstrained case, the low energy valence bands are twofold degenerate, whereas anisotropic strain lifts this degeneracy and moves the Dirac points away from the moir\'e Brillouin zone corners. The number of Dirac points is reduced from six in the unstrained case to four in the tilted two dimensional strained case and to two in the quasi one dimensional case. The white contours show equal energy lines. They preserve \(C_3\) symmetry in the unstrained case, while \(C_3\) symmetry is broken in both strained cases. In the quasi one dimensional case the contours are open and stripe like.}
    \label{fig:fig5}
\end{figure*}
\begin{table*}[t]
{ \caption{Geometric characterization of representative commensurate moir\'e supercells. Here $\textrm{p}_1,\textrm{p}_2$ are the lattice vector scaling factors and $\phi_1,\phi_2$ are the apparent rotation angles of the strained top layer primitive vectors relative to the bottom layer. $|\mathbf{A}_1|$ and $|\mathbf{A}_2|$ are the lengths of the moir\'e supercell lattice vectors ($\mathbf{A}_1=i\bm{a}'_1+j\bm{a}'_2=m\bm{a}_1+n\bm{a}_2$, $\mathbf{A}_2=k\bm{a}'_1+l\bm{a}'_2=q\bm{a}_1+r\bm{a}_2$), $\angle(\mathbf{A}_1,\mathbf{A}_2)$ is the angle between them, and $N_{\rm tot}(=2|il-jk|+2|mr-nq|)$ is the total number of atoms in the commensurate unit cell. The quasi 1D structure is finite and commensurate, but has a nearly rank one relative deformation matrix $S-I$, as quantified by $\eta\ll 1$.}
\label{tab:supercell}
\begin{ruledtabular}
\footnotesize
\begin{tabular}{c c c c c c c c c c}
Integers & Type
  & $\textrm{p}_1$ & $\textrm{p}_2$
  & $\phi_1$ & $\phi_2$
  & $|\mathbf{A}_1|$ & $|\mathbf{A}_2|$
  & $\angle(\mathbf{A}_1,\mathbf{A}_2)$
  & $N_{\rm tot}$ \\
\hline
$(11,5,6,11,11,6,5,11)$
  & Unstrained $6.008^\circ$
  & $1.0$ & $1.0$
  & $6.008^\circ$ & $6.008^\circ$
  & $23.47$\,\AA & $23.47$\,\AA
  & $60.0^\circ$ & $364$ \\
$(9,7,5,11,9,6,6,11)$
  & 2D tilted near $6.008^\circ$
  & $0.988$ & $1.015$
  & $-8.666^\circ$ & $-6.890^\circ$
  & $19.53$\,\AA & $23.47$\,\AA
  & $46.10^\circ$ & $254$ \\
$(10,5,9,10,11,6,10,11)$
  & Quasi 1D near $6.008^\circ$
  & $1.048$ & $1.009$
  & $\phantom{-}4.307^\circ$ & $-0.894^\circ$
  & $23.47$\,\AA & $25.92$\,\AA
  & $31.71^\circ$ & $232$ \\
\end{tabular}
\label{table3}
\end{ruledtabular}}
\end{table*}
In this section, we examine the effect of anisotropic strain on the electronic band structure of commensurate moir\'e solutions.
Fig.~\ref{fig:fig5} shows the band structure along high symmetry paths and the corresponding highest valence band over the first Brillouin zone for (Fig.~\ref{fig:fig5}(a,d)) the unstrained system at a twist angle of $\pm 6.008^\circ$, (Fig.~\ref{fig:fig5}(b,e)) a two dimensional (2D) strained commensurate solution, and (Fig.~\ref{fig:fig5}(c,f)) a quasi one dimensional (quasi 1D) commensurate solution. Upon introducing heterostrain, the layer exchange symmetry between the two graphene sheets is lifted. The four band Dirac structure of the unstrained system reorganizes such that only a two band Dirac like crossing remains near charge neutrality [Fig.~\ref{fig:fig5}(b)]. The states forming this crossing predominantly involve a single layer hybridized channel, while the remaining layer derived bands are shifted away from the Dirac energy and reconnect with the low energy manifold at higher symmetry points (notably $\Gamma$ and $M$). At these points, the larger little group symmetry constrains the allowed band representations and governs the symmetry allowed band reconnections. Breaking $C_3$ rotational symmetry while preserving time reversal symmetry releases Dirac points from being pinned to the high symmetry corners of the Brillouin zone, allowing them to move freely in momentum space and annihilate pairwise upon coincidence with oppositely charged partners. In a two dimensional lattice, however, the complete removal of all Dirac points is not generic, as simultaneous annihilation requires fine tuning of multiple parameters. Instead, the system naturally flows to the minimal configuration that remains generically stable in the absence of additional crystalline symmetries, consisting of two time reversal related pairs i.e., four Dirac points within the Brillouin zone [Fig.~\ref{fig:fig5}(e)]. Thus, heterostrain redistributes Dirac points in momentum space, splitting and displacing them without lifting their energy degeneracy or opening a gap \cite{Po}.
\begin{figure}
    \centering
    \includegraphics[width=\linewidth]{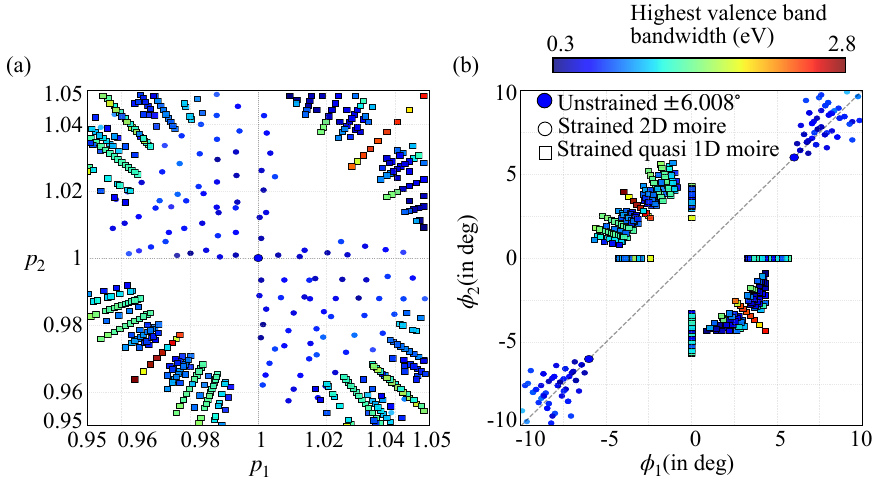}
    \caption{Highest valence band bandwidth \(W_v\) for the commensurate moir\'e solutions shown in Fig.~\ref{fig:fig2}, plotted in (a) the \((\textrm{p}_1,\textrm{p}_2)\) space and (b) the \((\phi_1,\phi_2)\) space. { The bandwidth is defined as
\(W_v=\max_{\mathbf k\in{\rm mBZ}}E_v(\mathbf k)-\min_{\mathbf k\in{\rm mBZ}}E_v(\mathbf k)\),
where \(E_v(\mathbf k)\) is the highest valence band over the first moir\'e Brillouin zone.} Large circles denote the unstrained reference configurations at \(\pm6.008^\circ\), small circles denote tilted two dimensional strained solutions, and squares denote quasi one dimensional strained solutions. Near the pristine reference points, tilted two dimensional solutions retain bandwidths comparable to the unstrained case, whereas quasi one dimensional solutions show a broader bandwidth distribution, reflecting stronger reconstruction of the low energy dispersion.}
    \label{fig:fig6}
\end{figure}
In the quasi one dimensional regime, strong moir\'e anisotropy effectively reduces the low energy electronic structure from two dimensions to one. One moir\'e reciprocal lattice vector becomes very small, producing a highly elongated Brillouin zone and strongly suppressing dispersion along the transverse momentum direction. As a result, the spectrum is dominated by channel like states dispersing along the stripe direction, with the transverse momentum acting only as a weak modulation. In this limit, band connectivity is no longer governed by symmetry constraints at isolated high symmetry points such as $\Gamma$ and $M$, as in the strained two dimensional case. Instead, band hybridization and reconnection occur at multiple momentum locations wherever the one dimensional dispersing bands intersect, reflecting the kinematically driven connectivity characteristic of quasi one dimensional systems [Fig.~\ref{fig:fig5}(c)]. This dimensional crossover also alters the stability criteria for Dirac crossings. While a two dimensional lattice with time reversal symmetry generically requires two time reversal related pairs (four Dirac points) for stability, the quasi one dimensional electronic structure is already stable with a single time reversal related pair [Fig.~\ref{fig:fig5}(f)]. As a consequence, only two Dirac points remain robust within the Brillouin zone in the quasi one dimensional regime, signalling a fundamental change in Dirac point topology driven by effective dimensional reduction.

Furthermore, in Fig.~\ref{fig:fig6} we show the bandwidth of the highest valence band of all the solutions in Fig.~\ref{fig:fig2}, in both the $(\textrm{p}_1, \textrm{p}_2)$ and $(\phi_1, \phi_2)$ parameter spaces. For the unstrained pristine twist angle of $\pm 6.008^\circ$, the highest valence band has a bandwidth of $0.885 eV$. The strained two dimensional (2D) moir\'e configurations exhibit bandwidths in the range $0.608$ – $1.296 eV$, with a mean value of $0.866 eV$. Notably, for strained 2D solutions close to the pristine configuration, the bandwidth remains comparable to the unstrained case in both parameter spaces, while larger bandwidths are predominantly associated with solutions farther from the pristine point. Although strain and tilt modify the shape of the Brillouin zone, the effective mass and coupling between moir\'e sites remain broadly comparable to those of the pristine system. 

In contrast, the quasi one dimensional (quasi 1D) moir\'e configurations display a much broader distribution of bandwidths, ranging from $0.3~ eV$ to $2.8~ eV$, with a mean value of $1.235~ eV$. We find that the low energy bandwidth of commensurate quasi one dimensional moir\'e configurations is primarily controlled by the separation between AA centres along the stripe direction, corresponding to the short axis of the moir\'e unit cell. Additional variations in the bandwidth arise from commensuration dependent stacking modulation, reflecting the phase sensitive nature of electronic dispersion. Along these effectively one dimensional paths, the kinetic energy along the stripe direction is less geometrically constrained than in two dimensional moir\'e configurations, potentially favouring an enhanced bandwidth when the AA-AA separation is small, but decreasing as the separation becomes large. { Thus, for tilted two dimensional strained solutions near the pristine configuration, the overall single particle dispersion scale remains close to that of the unstrained system, whereas quasi one dimensional solutions show a much stronger reconstruction of the low energy dispersion.}

\begin{figure}
    \centering
    \includegraphics[width=0.9\linewidth]{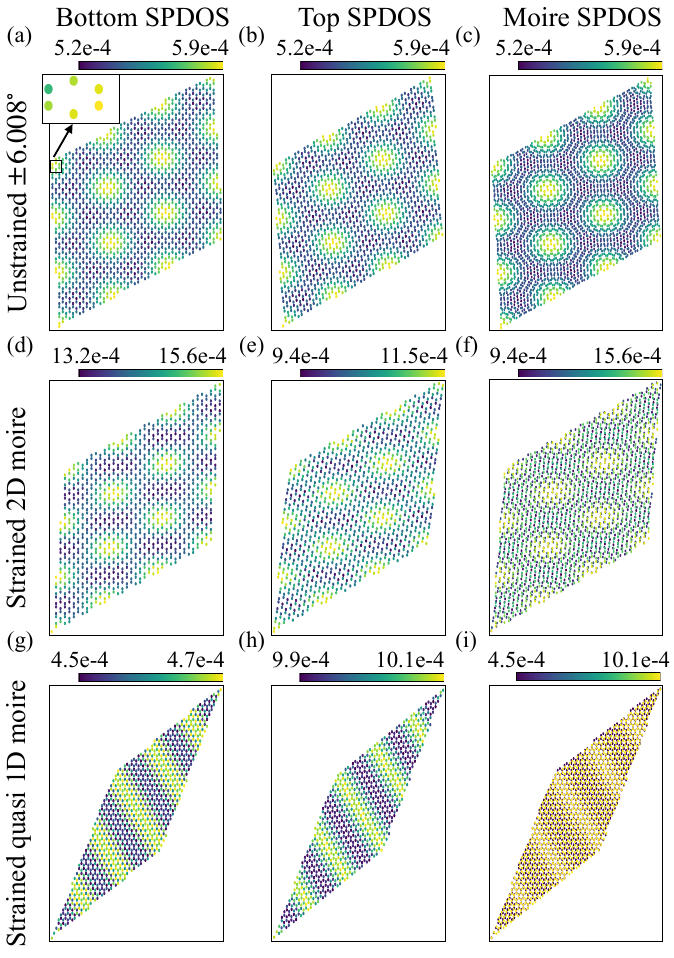}
    \caption{Spatially projected density of states (SPDOS) for the three representative commensurate structures described in table~\ref{table3}. { Panels (a-c) show the unstrained reference structure, panels (d-f) show the tilted two dimensional strained structure and panels (g-i) show the quasi one dimensional strained structure.} The unstrained and tilted two dimensional cases show spectral weight localized near AA regions, while the quasi one dimensional case shows enhanced stripe like localization.}
    \label{fig:fig7}
\end{figure}
\subsection{Spatially projected Density of States (SPDOS)}\label{SPDOS}
In this section, we visualize the distribution of the low energy electronic wave function in the presence of anisotropic strain for the three representative cases listed in Table~\ref{table3}, as shown in Fig.~\ref{fig:fig7}. Unlike the total density of states (DOS), which is averaged over the entire system, SPDOS provides information about how electronic states are distributed in real space. Mathematically, the spatial density of states at site $i$ is given by the diagonal elements of the imaginary part of the Green's function ~\cite{Souma}:
\begin{align}
 {\rm SPDOS}(K, r_i,E) = -\frac{1}{\pi}Im G_{ii}(K,E) \nonumber
\end{align}
where $G_{ii}(K,E)$ is the retarded Green's function at energy E. At the unstrained twist angle of $\pm 6.008^\circ$, the spatially resolved SPDOS at charge neutrality ($E=0$) is strongly modulated by the moir\'e pattern, with enhanced spectral weight localized at AA stacking regions (Fig.~\ref{fig:fig7}(a,b,c)). The absolute magnitude of the SPDOS at $E=0$ is, however, very small, reflecting the vanishing density of states of the dispersive Dirac bands at charge neutrality far from the flat band regime. The SPDOS contributions from the two graphene layers are identical in both spatial distribution and magnitude, consistent with the exact layer exchange symmetry of the unstrained system and the resulting layer degeneracy at the Dirac point. Within each layer, the SPDOS exhibits a sublattice dependent contrast originating from the pseudospin structure of Dirac eigenstates and its modulation by moir\'e induced interference, without implying any breaking of sublattice symmetry. At finite energies within the low energy Dirac bands $(E \ne 0)$, the SPDOS remains predominantly localized at AA stacking regions, although the localization contrast is reduced due to contributions from a broader range of momenta. The absolute SPDOS magnitude increases substantially away from $E=0$, reflecting the finite density of states of the dispersive Dirac bands.

In the two dimensional strained configuration, the low energy SPDOS at charge neutrality remains strongly localized at AA stacking regions, indicating that heterostrain does not suppress the real space moir\'e localization of Dirac states (Fig.~\ref{fig:fig7}(d,e,f)). While the spatial SPDOS patterns on the two layers remain identical, their magnitudes differ slightly. This modest difference in SPDOS intensity reflects an asymmetric distribution of the Dirac wavefunctions between the two layers. In the present configuration, the strained (top) layer satisfies $\textrm{p}_1 < 1, \textrm{p}_2 > 1$ and $|\textrm{p}_1-1|>|\textrm{p}_2-1|$, corresponding to an overall lattice compression and a consequent increase in intralayer hopping amplitudes. This leads to a slight broadening of the strained layer bands, which reduces their contribution to the low energy SPDOS. The magnitude and sign of this layer imbalance vary continuously with the strain parameters $\textrm{p}_1,\textrm{p}_2, \phi_1, \phi_2$.

In the quasi one dimensional strained configuration, the spatially resolved SPDOS exhibits a qualitative change compared to both the unstrained and two dimensional strained cases (Fig.~\ref{fig:fig7}(g,h,i)). Low energy electronic states preferentially accumulate along stripe like channels where the local stacking remains close to AA, leading to enhanced SPDOS intensity along these directions. As the quasi 1D moir\'e pattern emerges at larger values of $\textrm{p}_1$ and $\textrm{p}_2$, the top layer experiences substantially stronger lattice deformations, leading to pronounced asymmetry in the intralayer hopping amplitudes. Depending on whether the top layer is stretched or compressed relative to the bottom layer, the corresponding low energy bands become flatter or more dispersive, respectively. Since flatter bands contribute more strongly to the density of states, the low energy Dirac states acquire a larger weight on the layer with a smaller bandwidth. Consequently, a pronounced imbalance in the SPDOS magnitude develops between the two layers, while the spatial distribution of the SPDOS remains similar on both. For the configuration shown here, the top layer is stretched $(\textrm{p}_1,\textrm{p}_2>1)$, resulting in flatter low energy bands and a dominant contribution of the low energy electronic states from the top layer.
\begin{figure}
    \centering
    \includegraphics[width=0.85\linewidth]{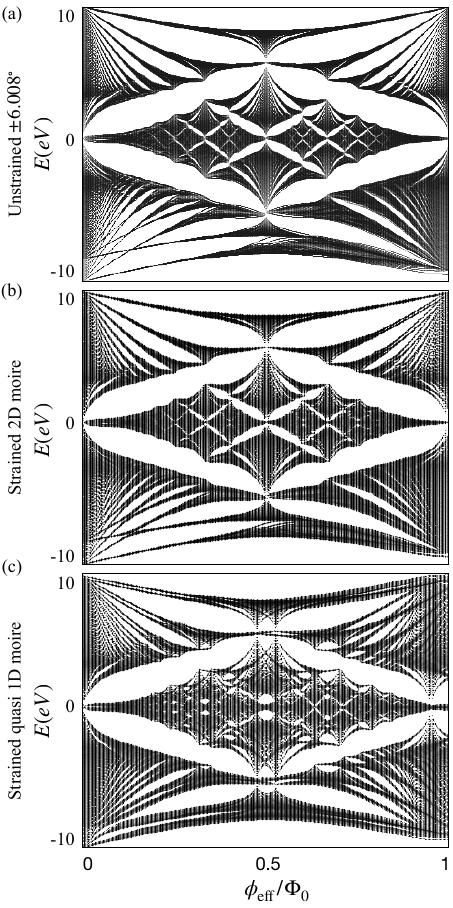}
    \caption{Hofstadter butterfly spectra for the representative commensurate structures listed in Table~\ref{table3}. The Peierls phase is implemented using the rational flux through the moir\'e supercell, \(\Phi_M/\Phi_0=x/y\), with \(\Phi_M=BA_M\). The horizontal axis is rescaled as an effective microscopic flux, \(\phi_{\rm eff}/\Phi_0=\Phi_M/[(N_{\rm tot}/4)\Phi_0]\), where \(N_{\rm tot}\) is the zero field atom count listed in Table~\ref{table3}. Panels (a), (b), and (c) correspond to the unstrained, tilted two dimensional strained, and quasi one dimensional strained structures, respectively. The quasi one dimensional case shows stronger Hofstadter splitting than the tilted two dimensional case.}
    \label{fig:fig8}
\end{figure}

\subsection{Hofstadter Butterfly under anisotropic strain}\label{Butterfly}
{ In this section, we compute the Hofstadter spectra using the periodic Landau gauge Peierls substitution, since the strained commensurate moir\'e cells are non orthogonal and include long range intra and interlayer hoppings~\cite{Hasegawa}. The rational flux entering the calculation is the flux through the moir\'e supercell, $ \frac{\Phi_M}{\Phi_0}=\frac{x}{y}, \Phi_M=BA_M , $ where \(A_M\) is the moir\'e unit cell area and \(\Phi_0=h/e\). For Fig.~\ref{fig:fig8}, we rescale the horizontal axis to an effective microscopic flux, $ \frac{\phi_{\rm eff}}{\Phi_0} = \frac{\Phi_M}{(N_{\rm tot}/4)\Phi_0}, $ where \(N_{\rm tot}\) is the number of atoms in the zero field commensurate moir\'e unit cell, listed for each structure in Table~\ref{table3}. This rescaling puts the unstrained, tilted two dimensional strained, and quasi one dimensional strained structures on the same effective graphene cell flux scale, allowing their Hofstadter spectra to be compared despite their different moir\'e unit cell areas and atom counts. Details of the Peierls phases, magnetic Bloch Hamiltonian, and flux convention are given in Appendix~\ref{app:hofstadter_method}.} In the unstrained (pristine) Hofstadter butterfly, Landau level degeneracy is protected by the high symmetry of the band structure. Multiple symmetry related valley pockets are exactly degenerate and well separated in momentum space (Fig.~\ref{fig:fig5}(d)), so semiclassical cyclotron orbits remain independent, and Landau quantization yields a degenerate spectrum with the familiar half integer Hall sequence (Fig.~\ref{fig:fig8}(a)) \cite{Hasegawa}.
\par Introducing two dimensional anisotropic strain explicitly breaks \(C_3\) symmetry, but does not qualitatively alter this picture over the low flux range shown in Fig.~\ref{fig:fig8}(b). In particular, no appreciable splitting is observed up to \(\phi_{\rm eff}/\Phi_0\simeq0.3\), corresponding to an extremely large microscopic field scale of order \(10^4\)~T. In the 2D strained case, four valley pockets persist inside the Brillouin zone and remain degenerate and well isolated up to a Lifshitz energy ($E_L$), as revealed by the equal energy contours (Fig.~\ref{fig:fig5}(e)). Below $E_L$, Landau levels form independently within each closed valley orbit, and no resolvable splitting appears despite the reduced symmetry. Only above the Lifshitz transition do the pockets merge through narrow saddle point necks, enabling inter valley coupling in principle. However, the resulting splitting is controlled by magnetic breakdown across these localized saddle regions, with a tunnelling probability $P_{MB} \sim exp(-B_0/B)$, where $B_0$ is the magnetic breakdown field and is determined by the saddle point energy gap and local curvature \cite{Blount, Shoenberg}. Because the necks are sharp and the Lifshitz energy lies relatively high, $B_0$ is large, leading to an exponential suppression of tunnelling at experimentally relevant fields. Consequently, valley hybridization and Landau level splitting occur only at extremely high magnetic fields, rendering the low field butterfly essentially indistinguishable from the pristine case within a single particle description. 
In stark contrast, the quasi one dimensional strained system exhibits immediate butterfly splitting for any nonzero magnetic field (Fig.~\ref{fig:fig8}(c)). Here, only two valley pockets survive, and while the dispersion near each valley remains approximately isotropic at low energies, the band becomes nearly flat along the strained direction away from the valley centres, causing the equal energy contours to open into extended trajectories spanning large regions of the Brillouin zone (Fig.~\ref{fig:fig5}(f)). As a result, semiclassical orbits associated with different valleys are no longer separated by localized saddle points but remain weakly separated over extended momentum space segments. This invalidates the standard magnetic breakdown picture based on isolated tunnelling events: hybridization is accumulated continuously along the orbit rather than being triggered at a single bottleneck. The lifting of Landau level degeneracy therefore occurs already at an infinitesimal magnetic field, producing two distinct Hofstadter butterflies with different periodicities. Thus, while splitting in the pristine and 2D strained cases is suppressed by symmetry and exponentially weak magnetic breakdown, respectively, quasi one dimensional strain reshapes the semiclassical phase space connectivity, leading to an immediate and robust splitting mechanism.
{ That two dimensional and quasi one dimensional moir\'e superlattices yield different Hofstadter spectra is expected once a magnetic length is introduced. The key result is that the degeneracy lifting mechanism in the quasi one dimensional regime is qualitatively distinct from the magnetic breakdown picture governing the two dimensional strained case: instead of exponentially suppressed tunnelling at isolated saddle points, the quasi one dimensional regime allows continuous hybridization along extended momentum space segments. This mechanism cannot be inferred from the two dimensional strained case alone.}
\subsection{Anisotropic strain near magic angle TBG}\label{magic}
\begin{figure}
    \centering
    \includegraphics[width=0.9\linewidth]{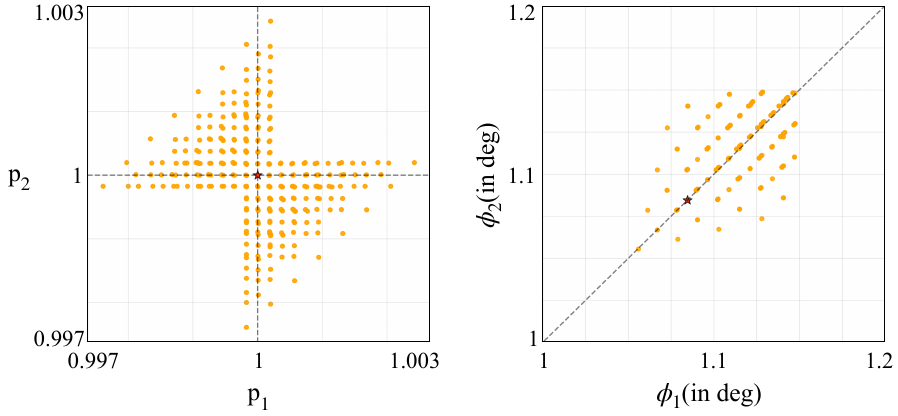}
    \caption{Commensurate solutions near the magic angle reference twist \(\theta_0\sim 1.084^\circ\), shown in the (left) \((\textrm{p}_1,\textrm{p}_2)\) and (right) \((\phi_1,\phi_2)\) parameter spaces. Here \(\textrm{p}_i\) denotes the scaling of the \(i\)th strained primitive lattice vector, and \(\phi_i\) denotes its apparent rotation relative to the corresponding bottom layer primitive vector. The dashed lines mark \(\textrm{p}_1=\textrm{p}_2=1\) in the left panel and \(\phi_1=\phi_2\) in the right panel, while the star indicates the near pristine reference configuration. For clarity, only solutions within a narrow window of lattice vector strain, \(|\textrm{p}_i-1|\lesssim0.003\), and apparent rotation near the magic angle are shown.}
    \label{fig:fig9}
\end{figure}

\begin{figure}
    \centering
    \includegraphics[width=0.9\linewidth]{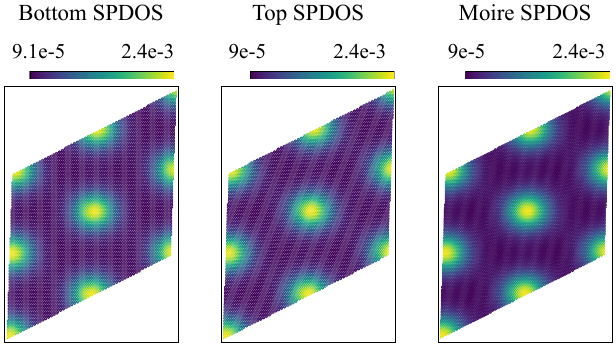}
    \caption{Spatially projected density of states (SPDOS) for the representative weakly strained commensurate configuration near the magic angle regime. The left and middle panels show the layer resolved SPDOS on the bottom and top layers, respectively, while the right panel shows the total moir\'e resolved SPDOS. The low energy spectral weight remains concentrated near the triangular network of AA-stacked regions in both layers, showing that this weak residual anisotropic strain does not destroy the real space AA localization characteristic of the magic angle regime.}
    \label{fig:fig10}
\end{figure}

\begin{figure}
    \centering
    \includegraphics[width=0.9\linewidth]{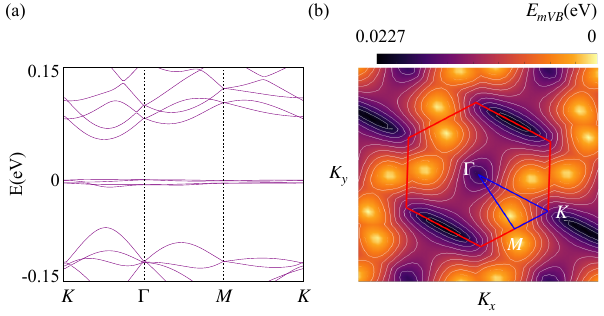}
    \caption{(a) Band structure along the high symmetry path for the representative strained commensurate configuration near the twist angle $\sim 1.084 ^\circ$, (b) two dimensional surface plot of the maximum valence band energy $E_{mVB}$ in the first moiré Brillouin zone for the same solution. As in other two dimensional moiré systems, four Dirac points are clearly visible within the Brillouin zone. The bandwidth of the lowest energy valence band is approximately $22.7$meV, which is comparable to the characteristic low energy bandwidth associated with the magic angle regime.}
    \label{fig:fig11}
\end{figure}

In this section, we systematically search for two dimensional anisotropically strained commensurate solutions in the vicinity of the magic angle $\sim 1.084^\circ$. { The representative solution discussed in this section is not intended to illustrate a strong anisotropic strain effect. Rather, it is a weakly strained, nearly pristine commensurate solution within the experimentally relevant  distortion window, used to test whether the tilted two dimensional regime identified above retains magic angle like features when the twist angle is close to \(1.084^\circ\).} The unstrained magic angle configuration ($n_0,m_0 = 30,31$) is characterized by the integer set ({\it i, j, k, l, m, n, q, r}) = (61, 30, 31, 61, 61, 31, 30, 61). As the search range of integers increases for smaller twist angles, the number of admissible commensurate solutions grows rapidly. In Fig.~\ref{fig:fig9}, we display only those solutions confined within a narrow window of strain and relative rotation, focusing exclusively on 2D moiré configurations. For a representative solution 
($\textrm{p}_1=1.000397,\textrm{p}_2=0.999613,\phi_1=1.148^\circ, \phi_2 = 1.147^\circ$ , corresponding to the eight integer set $31,59,59,30,30,59,59,31$), we examine the spatially projected density of states (Fig.~\ref{fig:fig10}) and the maximum valence band dispersion in the first Brillouin zone (Fig.~\ref{fig:fig11}). The low energy electronic states remain strongly localized in the AA-stacked regions, consistent with the magic angle behaviour. Moreover, four Dirac points are present within the first Brillouin zone, as expected for a strained 2D moiré pattern, and the bandwidth of the lowest energy band is approximately $22.7$ meV, comparable to that of the magic angle regime \cite{Mondal}. These results support our claim that multiple commensurate solutions exist within the experimentally observed distortion range, exhibiting similar localization characteristics, Dirac topology, and narrow bandwidth.

\section{Conclusions}\label{conclu}
In this work, we investigated the origin of magic angle phenomenology in twisted bilayer graphene over a finite range of twist angles and anisotropic strain, motivated by growing experimental evidence that correlated physics is not restricted to a single pristine geometry. Focusing on commensurate moir\'e superlattices generated by a fixed moir\'e deformation across different twist angles, with particular emphasis on the $\pm 6.008^\circ$ case, we systematically classified the allowed strained configurations and analyzed their electronic and magnetic field responses. We demonstrated that two distinct classes of commensurate moir\'e patterns can emerge near a pristine twist angle depending on the relative rotation of the strained lattice vectors: tilted two dimensional moir\'e superlattices and quasi one dimensional stripe like moir\'e patterns. Anisotropic strain strongly reshapes the band structure and reduces the number of Dirac points within the moir\'e Brillouin zone, leading to qualitatively different low energy physics in these two regimes.
A central finding of our study is the robustness of commensurate two dimensional moir\'e patterns near the pristine twist angle. Despite broken rotational and layer symmetries, these tilted two dimensional structures retain triangular AA stacked regions with bandwidths comparable to the unstrained case, supporting similar single particle dispersion scales over a finite strain and twist angle window. Electron localization at AA regions remains robust against layer asymmetry, and up to the highest experimentally accessible magnetic fields, the Landau level structure closely resembles that of the neighbouring unstrained system. These results provide a natural explanation for the persistence of magic angle phenomenology across a finite distortion range. In contrast, quasi one dimensional moir\'e patterns arising from different strain rotation configurations exhibit a qualitatively distinct behavior. The effective dimensional reduction leads to strong band structure reorganization, layer polarized electronic states, and immediate hybridization of Landau levels, resulting in pronounced splitting of the Hofstadter spectrum even at infinitesimal magnetic fields. These extrinsic features suggest that quasi one dimensional strained moir\'e systems host rich physics beyond the scope of conventional magic angle behaviour and warrant further investigation.
Overall, our results establish anisotropic strain not as a source of disorder, but as a powerful geometric control parameter that organizes commensurate moir\'e structures. By identifying the conditions under which magic angle physics is robust versus fundamentally altered, this work provides a unified framework for understanding experimental variability and offers a roadmap for the deliberate engineering of strained moir\'e materials.
\section*{ACKNOWLEDGMENTS}
The authors acknowledge the support provided by the KEPLER computing facility, maintained by the Department of Physical Sciences, IISER Kolkata. A.M. acknowledge financial support from IISER Kolkata through the Institute PhD Fellowship. B.L.C. acknowledges the SERB for Grant No. SRG/2022/001102 and “IISER Kolkata Start-up-Grant” Ref. No. IISER-K/DoRD/SUG/BC/2021-22/376.
\appendix
\renewcommand\thefigure{\thesection} 
\setcounter{figure}{0}
{ \section{Mapping the eight integer construction to physical strain parameters}
\label{appA}

In this appendix we provide the explicit steps used to convert each eight integer commensurate solution ($i,j,k,l,m,n,q,r$) into the lattice vector parameters \((\textrm{p}_1,\textrm{p}_2,\phi_1,\phi_2)\) and into physically transparent strain variables. This makes the commensurate search and the strain classification reproducible. 
The common moir\'e supercell is defined by
\[
\begin{pmatrix}
\mathbf A_1\\
\mathbf A_2
\end{pmatrix}
=
\begin{pmatrix}
i & j\\
k & l
\end{pmatrix}
\begin{pmatrix}
\bm{a}_1^\prime\\
\bm{a}_2^\prime
\end{pmatrix}
=
\begin{pmatrix}
m & n\\
q & r
\end{pmatrix}
\begin{pmatrix}
\bm{a}_1\\
\bm{a}_2
\end{pmatrix},
\]
where \(\bm{a}_1,\bm{a}_2\) are the unstrained bottom layer primitive vectors and \(\bm{a}_1^\prime,\bm{a}_2^\prime\) are the strained top layer primitive vectors. Solving for the strained primitive vectors gives
\[
\begin{pmatrix}
\bm{a}_1^\prime\\
\bm{a}_2^\prime
\end{pmatrix}
=
\frac{1}{il-jk}
\begin{pmatrix}
lm-jq & ln-jr\\
-km+iq & -kn+ir
\end{pmatrix}
\begin{pmatrix}
\bm{a}_1\\
\bm{a}_2
\end{pmatrix}.
\]
Thus we may write
\[
\bm{a}_1^\prime=e \bm{a}_1+f\bm{a}_2,\qquad
\bm{a}_2^\prime=g\bm{a}_1+h\bm{a}_2,
\]
with
\begin{align}
e &= \frac{lm-jq}{il-jk}, &
f &= \frac{ln-jr}{il-jk}, \nonumber\\
g &= \frac{-km+iq}{il-jk}, &
h &= \frac{-kn+ir}{il-jk}.
\end{align}
Using the graphene primitive vectors $
\bm{a}_1=a(1,0),
\bm{a}_2=a\left(-\frac12,\frac{\sqrt3}{2}\right),
$
the lattice vector scaling factors and apparent rotations are
\begin{align}
\textrm{p}_1 &= \sqrt{e^2+f^2-ef}, \nonumber\\
\phi_1 &= \operatorname{atan2}\left(f\sqrt3,\,2e-f\right), \nonumber\\
\textrm{p}_2 &= \sqrt{g^2+h^2-gh}, \nonumber\\
\phi_2 &= \operatorname{atan2}\left(g\sqrt3,\,g-2h\right).
\end{align}
Here \(\operatorname{atan2}\) is used to fix the quadrant of the rotation angle. The parameters \(\textrm{p}_1,\textrm{p}_2,\phi_1,\phi_2\) provide a convenient lattice vector parametrization of the strained top layer. To obtain the physical strain associated with the same commensurate solution, we construct the Cartesian deformation matrix
\[
S=(\bm{a}'_1\ \bm{a}'_2)(\bm{a}_1\ \bm{a}_2)^{-1}.
\]
This matrix represents one homogeneous deformation of the top layer relative to the unstrained bottom layer. We separate the rigid rotation from the elastic deformation using the polar decomposition
\[
S=WP=WVDV^T,
\]
where \(W\) is a rotation matrix, \(P\) is a symmetric positive definite stretch matrix, \(V\) contains the principal stretching directions, and
\[
D=\mathrm{diag}(d_1,d_2),\qquad d_1\ge d_2 .
\]
The rotation angle \(\xi\) extracted from \(W\) gives the overall rigid rotation of the top layer. The finite strain tensor is
\[
\epsilon=P-I.
\]
The principal strains are $ \epsilon_1=d_1-1, \epsilon_2=d_2-1, $
the strain anisotropy is $ \kappa=\frac{d_1}{d_2},$ and the principal strain direction \(\psi\) is obtained from the eigenvector of \(P\) corresponding to \(d_1\). The area change is
\[
\frac{\Delta A}{A}= |S|-1=d_1d_2-1.
\]
These quantities translate each eight integer solution into physical strain parameters.
\section{Apparent rotations of primitive lattice vectors}
\label{app:apparent_rotation}

Here we explain why the two primitive lattice vectors of graphene can acquire different apparent rotations even though they are generated by a single homogeneous deformation matrix \(S\).
For a unit vector initially oriented at an angle \(\alpha\),
\[
\hat{\mathbf n}(\alpha)=(\cos\alpha,\sin\alpha)^T ,
\]
the apparent angular change after deformation is
\begin{equation}
\Delta\theta(\alpha)
=
\arg\!\left[S\hat{\mathbf n}(\alpha)\right]-\alpha .
\label{eq:app_exact_delta_theta}
\end{equation}
Using the polar decomposition \(S=WP\), this separates into
\begin{equation}
\Delta\theta(\alpha)=\xi+\delta\theta_P(\alpha),
\label{eq:app_delta_split}
\end{equation}
where \(\xi\) is the common rigid rotation from \(W\), and \(\delta\theta_P(\alpha)\) is the direction dependent angular change produced by the stretch matrix \(P\).

Writing
\[
P=
R_\psi
\begin{pmatrix}
d_1 & 0\\
0 & d_2
\end{pmatrix}
R_\psi^T,
\qquad d_1\ge d_2 ,
\]
and defining \(\beta=\alpha-\psi\), the stretch changes the direction
\((\cos\beta,\sin\beta)^T\) into \((d_1\cos\beta,d_2\sin\beta)^T\). Hence
\begin{equation}
\begin{aligned}
\delta\theta_P(\alpha)
&=
\operatorname{atan2}
\left[
d_2\sin(\alpha-\psi),
d_1\cos(\alpha-\psi)
\right]
-(\alpha-\psi).
\end{aligned}
\label{eq:app_delta_theta_P}
\end{equation}
This expression is exact for any nonsingular homogeneous deformation and shows explicitly that the stretch induced angular change depends on the initial direction \(\alpha\).

For graphene, $
\bm{a}_1=a(1,0), \bm{a}_2=a\left(-\frac12,\frac{\sqrt3}{2}\right),$
so that
$
\alpha_1=0^\circ, \alpha_2=120^\circ .
$
The apparent rotations of the two primitive vectors are therefore
\begin{equation}
\phi_i=\Delta\theta(\alpha_i)
=
\xi+\delta\theta_P(\alpha_i),
\qquad i=1,2 .
\label{eq:app_phi_i}
\end{equation}
Since \(\alpha_1\neq\alpha_2\), the two stretch induced contributions are generally different. Thus a single homogeneous anisotropic strain can rotate the two graphene primitive lattice directions by different amounts.

The sign of the stretch induced contribution follows from Eq.~\eqref{eq:app_delta_theta_P}. Using the tangent subtraction formula,
\begin{equation}
\tan\delta\theta_P(\alpha)
=
-\frac{(d_1-d_2)\sin[2(\alpha-\psi)]}
{2\left[d_1\cos^2(\alpha-\psi)+d_2\sin^2(\alpha-\psi)\right]} .
\label{eq:app_sign_delta_theta}
\end{equation}
The denominator is positive and \(d_1\ge d_2\), so the sign is controlled by $ \sin[2(\alpha-\psi)] .$
Since the graphene primitive vectors are separated by \(120^\circ\), the factors $ \sin[2(\alpha_1-\psi)]$ and $\sin[2(\alpha_2-\psi)] $ can have different signs depending on the principal strain direction \(\psi\). Therefore the same anisotropic stretch can generate opposite stretch induced angular changes for the two primitive vectors.

The full apparent rotations also contain the common rigid rotation \(\xi\). Hence, even if the stretch induced parts have opposite signs, \(\phi_1\) and \(\phi_2\) can still have the same sign when \(\xi\) dominates, as in tilted two dimensional solutions. In contrast, in quasi one dimensional solutions, the anisotropic or shear contribution can be large enough that the full apparent rotations acquire opposite signs. As graphene primitive vectors are separated by \(120^\circ\), the same shear containing deformation can rotate them by different amounts. Thus \(\phi_1\) and \(\phi_2\) are not independent rotations, but both follow from the same deformation matrix \(S\).

\section{Hofstadter butterfly calculation}
\label{app:hofstadter_method}
In this appendix, we summarize the numerical procedure used to obtain the Hofstadter spectra in Fig.~\ref{fig:fig8}. For each commensurate structure, the moir\'e lattice vectors are
\begin{align}
\mathbf A_1 &= i\bm{a}'_1+j\bm{a}'_2
            = m\bm{a}_1+n\bm{a}_2,\\
\mathbf A_2 &= k\bm{a}'_1+l\bm{a}'_2
            = q\bm{a}_1+r\bm{a}_2 .
\end{align}
The moir\'e unit cell area is
\(A_M=|\mathbf A_1\times\mathbf A_2|\), and the rational flux entering the calculation is the moir\'e supercell flux
\[
\frac{\Phi_M}{\Phi_0}=\frac{x}{y},
\qquad
\Phi_M=BA_M,
\qquad
\Phi_0=\frac{h}{e}.
\]
In the plotted spectra, this flux is rescaled to the effective microscopic flux
\[
\frac{\phi_{\rm eff}}{\Phi_0}
=
\frac{\Phi_M}{(N_{\rm tot}/4)\Phi_0},
\]
where \(N_{\rm tot}\) is the number of atoms in the zero field commensurate moir\'e unit cell.
The magnetic field is included through the Peierls substitution,
\[
t_{ij}\rightarrow t_{ij}\exp(i\theta_{ij}),
\]
where $
\theta_{ij}
=
\frac{2\pi}{\Phi_0}
\int_{\mathbf r_j}^{\mathbf r_i}
\mathbf A(\mathbf r)\cdot d\mathbf l .
$
Here \(t_{ij}\) is the Slater Koster hopping used in the zero field tight binding calculation.

As the strained moir\'e cells are generally non orthogonal, we use a periodic Landau gauge written in oblique moir\'e coordinates. We define
\[
\mathbf r=\xi_1\mathbf A_1+\xi_2\mathbf A_2,
\]
and reciprocal vectors \(\mathbf G_1^M,\mathbf G_2^M\) satisfying
\[
\mathbf G_\mu^M\cdot\mathbf A_\nu=2\pi\delta_{\mu\nu}.
\]
The vector potential is chosen as
\begin{equation}
\begin{aligned}
\mathbf A_{\rm pL}(\mathbf r)
&=
\frac{B A_M}{2\pi}
\Bigg[
(\xi_1-\lfloor \xi_1\rfloor)\mathbf G_2^M  \\
&\qquad
-
\xi_2
\sum_{N=-\infty}^{\infty}
\delta(\xi_1-N+\epsilon)\mathbf G_1^M
\Bigg],
\end{aligned}
\label{eq:periodic_landau_gauge}
\end{equation}
where \(\epsilon\) is a positive infinitesimal fixing the discontinuity. This gauge is periodic along \(\mathbf A_1\).
For rational moir\'e-supercell flux
\[
\frac{\Phi_M}{\Phi_0}=\frac{x}{y},
\]
with \(x\) and \(y\) coprime, the magnetic periodicity is restored by enlarging the unit cell along \(\mathbf A_2\):
\[
\mathbf A^{\rm mag}_1=\mathbf A_1,\qquad
\mathbf A^{\rm mag}_2=y\mathbf A_2 .
\]
The effective microscopic flux \(\phi_{\rm eff}/\Phi_0\) is used only as the rescaled horizontal axis in Fig.~\ref{fig:fig8}. The magnetic Bloch Hamiltonian is then constructed as
\[
H_{\alpha\beta}(\mathbf k;B)
=
\sum_{\mathbf R}
t_{\alpha\beta}(\mathbf R)
e^{i\theta_{\alpha\beta}(\mathbf R)}
e^{i\mathbf k\cdot\mathbf R},
\]
where \(\alpha,\beta\) label orbitals in the magnetic unit cell and
\[
\mathbf R=n_1\mathbf A_1^{\rm mag}+n_2\mathbf A_2^{\rm mag}.
\]
The Peierls phase \(\theta_{\alpha\beta}(\mathbf R)\) is evaluated by integrating Eq.~\eqref{eq:periodic_landau_gauge} along the straight hopping path. If the path crosses a discontinuity line \(\xi_1=N\), the delta function term in Eq.~\eqref{eq:periodic_landau_gauge} is included. This ensures that the phase accumulated around a closed loop satisfies
\[
\sum_{\rm loop}\theta_{ij}
=
2\pi\frac{B A_{\rm loop}}{\Phi_0},
\]
where \(A_{\rm loop}\) is the signed area enclosed by the loop.
For each rational value of \(\Phi_M/\Phi_0\), we diagonalize \(H(\mathbf k;B)\) and collect the eigenvalues. Repeating this over the chosen flux values gives the Hofstadter butterfly.}

\end{document}